# Selective incorporation of antimony into gallium nitride

Yujie Liu,[1] Ishtiaque Ahmed Navid,[2] Zetian Mi,[2] and Emmanouil Kioupakis[1,a)]

[1]*Department of Materials Science and Engineering, University of Michigan, Ann Arbor, Michigan 48109, USA*

[2]*Department of Electrical Engineering and Computer Science, University of Michigan, Ann Arbor, 1301 Beal Avenue, Ann Arbor, Michigan 48109, USA*

**Dilute concentrations of antimony (Sb) incorporation into GaN induce strong band-gap bowing and tunable room-temperature photoluminescence from the UV to the green spectral regions. However, the atomistic details of the incorporation of Sb into the GaN host remain unclear. In this work, we use first-principles calculations to understand the thermodynamics of Sb substitution into GaN, and its effect on the optical and Raman spectra. Although it is empirically considered that Sb is preferentially incorporated as an anion (Sb$^{3-}$) into the N sublattice, we demonstrate that Sb can also be incorporated as a cation (Sb$^{3+}$, Sb$^{5+}$) into the metal sublattice. Our thermodynamic analysis demonstrates that Sb$_N$[9], Sb$_{Ga}$[2+], and Sb$_{Ga}$[9] can co-exist under Ga-rich conditions in n-type samples. We further confirm the dual incorporation of Sb by calculating the vibrational frequencies of different anionic and cation substitutions to explain the origins of experimentally observed additional Raman peaks of Sb-doped GaN. Moreover, the calculated band structures of different Sb substitutions into GaN explain the experimental photoluminescence and optical absorption spectra. Overall, our analysis suggests that the coexistence of Sb$^{3-}$, Sb$^{3+}$, and Sb$^{5+}$ substitutions into GaN explains the totality of experimental measurements. Our results demonstrate that the selective incorporation of Sb into GaN (and potentially other group-V elements such as As, P, or Bi) by tuning the growth conditions can drastically modify the electronic properties, for applications in visible light emitters and photocatalysis.**

Gallium nitride (GaN) is a wide-band-gap semiconducting material that is widely deployed in electronic and optoelectronic applications due to its outstanding properties, including efficient light emission, high electron mobility, excellent thermal stability, and high breakdown voltage.[1–3] These characteristics make it an extremely promising material for a broad range of applications in solid-state lighting, telecommunications, and energy conversion.[4–7] However, although the alloys of GaN are widely used in blue and green LEDs and laser diodes such as InGaN,[8] their applications in the long-wavelength part of the visible spectrum and in the infrared (IR) region are constrained by their wide band gap and difficulties in achieving high indium concentrations in GaN.

Highly mismatched alloys of GaN that incorporate dilute concentrations of elements with distinctly different properties (e.g., ionic size, electronegativity, etc.) such as Sb are a promising approach to drastically modify the band structure of GaN and reduce its band gap for applications in the visible and the mid-IR.[9] In particular, dilute-antimonide III-nitride alloys may enable high-performance optoelectronic devices that operate in the IR region, which is critical for a wide range of applications such as chemical sensing, remote sensing, and communications.[10–23] Previous experimental[10,12,13,15,17,19,21,22] and computational

a) Author to whom correspondence should be addressed: kioup@umich.edu

studies[24-27] have demonstrated that the incorporation of dilute concentrations (on the order of 1%) of Sb in GaN can lead to an abrupt reduction of the band gap from 3.4 eV to ~2 eV. This characteristic makes it a promising candidate for visible optoelectronics devices as it exhibits strong emission in the deep-visible spectral range, and also makes it suitable for solar water splitting and nitrogen fixation photocatalysis.[28]

However, there is a lack of thorough theoretical explanations for the mechanism behind the observed features emerging in Sb-doped GaN experiments, including large differences in the band gap values obtained from photoluminescence (PL) (~2.0 eV),[12,19] optical absorption spectra (~1.0 eV),[14] transmission measurements (~1.9 eV),[13] and the appearance of new peaks in micro-Raman spectra.[21] The sources of these experimental features cannot be explained under the assumption that Sb is incorporated as a substitution in the anion sites of GaN ($Sb_N$), indicating that other incorporation sites may also be possible. In previous work by Desnica $et\ al.$, the term "antisite" incorporation of Sb, i.e., on the metal site of GaN, has been used to explain the experimental Mössbauer spectra of Sb-implanted GaN.[29] All these unsolved mysteries underscore the necessity for computational analysis utilizing density functional theory (DFT). Previous computational studies assumed anion substitution to evaluate the band structures, both from a first-principles perspective[24,25] and within the band-anticrossing (BAC) model.[26,27] However, the BAC model lacks atomistic understandings of, e.g., the Sb orbitals that contribute to the band-edge states. Moreover, although a previous DFT study[25] explored the contribution of Sb orbitals to edge states from a first-principles perspective, it did so only within the context of anion substitution. At the same time, optical calculations for the Raman signatures of Sb-doped GaN remain unexplored, and the substitutions of Sb at cation sites ($Sb_{Ga}$) have not been thoroughly investigated. Additionally, other prevalent group-V dopants, such as arsenic (As) and bismuth (Bi) have demonstrated a stronger thermodynamic propensity to act as cations when introduced into GaN,[30,31] which suggests a systematic trend among all these group-V dopants. Collectively, the totality of experimental and computational findings suggests that Sb substituting exclusively on the anion is insufficient to elucidate its role in GaN, underscoring the need for a more systematic understanding of Sb incorporation into GaN.

In this work, we apply first-principles computational methods based on hybrid DFT to examine the incorporation of dilute Sb into GaN and understand its effect on the electronic and optical properties. Our calculations for the formation energies of substitutional Sb impurities in bulk GaN find that both anionic Sb ($Sb_N$) and cationic Sb ($Sb_{Ga}$) are stable substitutions under different growth conditions. Moreover, the calculated phonon frequencies for different charge states and Sb incorporation sites explain the origin of the additional Sb-related peaks in the measured Raman spectra of GaSbN. Additionally, our calculated band structures verify that anionic $Sb_N$ and cationic $Sb_{Ga}$ can explain the observed PL and optical absorption measurements, respectively. The combination of our theoretical findings with experimental data uncovers the selective incorporation of Sb into dilute GaSbN alloys under different growth conditions, provides insights on the effect of substitutional Sb on the GaN band structure, and demonstrates the potential of GaSbN alloys for applications in optoelectronic and photonic GaN-based devices in the visible.

Our atomistic calculations are based on density functional theory with the Heyd, Scuseria, and Ernzerhof (HSE) hybrid exchange-correlation functional[32] as implemented in VASP.[33] To ensure sufficient convergence of the dilute dopant properties at a reasonable computational cost, we use a 96-atom supercell of GaSbN with one Sb substitutional atom in the simulation cell, corresponding to an atomic concentration of ~2% in order to remain in the dilute limit. A plane-wave energy cutoff of 400 eV and a $2 \times 2 \times 2$ Monkhorst-Pack Brillouin-zone sampling mesh are used, along with projector-augmented wave (PAW) pseudopotentials.[34] Further details regarding the crystal structure of the supercell can be found in Supplementary Material. We set the fraction of the exact exchange, which is the mixing parameter in HSE, to 0.3 in order to recover the experimental band gap (~3.44 eV) of binary GaN. We follow the established computational approach for the calculation of defect formation energies and transition levels.[35] The formation energy of, e.g., $Sb_N$ in charge state $q$ is given by

$$E^f\left(Sb_N^q\right) = E^{tot}\left(Sb_N^q\right) - E^{tot}(GaN_{bulk}) + \sum n_i \mu_i + q(E_F + E_V) + \Delta E_{corr}\left(Sb_N^q\right), \tag{1}$$

where $E^{tot}\left(Sb_N^q\right)$ is the total energy of defective GaN containing a substitutional Sb on an N site after structural relaxation, and $E^{tot}(GaN_{bulk})$ is the total energy of pristine bulk GaN for the same supercell size. The $\sum n_i \mu_i$ term represents the summation of the number of atoms involved in forming the defect times their corresponding chemical potential. As an example, for $Sb_N$, this term is $\mu_N - \mu_{Ga}$. The Fermi energy $E_F$ in the charge term is referenced to the valence-band maximum (VBM) $E_V$. The last correction term accounts for finite-size effects of charged defects following methods based on the analysis of the electrostatic potential as established in previous works.[36,37] For Raman spectra calculations, we utilize density functional perturbation theory (DFPT)[38] as implemented in Quantum ESPRESSO[39,40] to calculate the Raman intensities of Raman-active phonon modes for various Sb substitutions into GaN and explore the origins of the experimentally observed peaks. We utilize the Perdew-Wang[41] local density approximation (LDA) exchange–correlation functional with optimized norm-conserving Vanderbilt pseudopotentials[42] obtained by PseudoDojo.[43] A wave-function cutoff of 120 Ry is chosen, and the self-consistency threshold is set to $10^{-13}$ $Ry^2$ for the phonon calculations.

Although the possibility of formations of substitutional Sb cations has not been as widely discussed as anionic substitutions, we can understand its origin based on the ionic radius and electronegativity of Sb. First, the electronegativity of Sb (1.9) lies between that of Ga (1.6) and N (3.0),[44] indicating that it can either act as an anion replacing N, as in, e.g., the GaSb

III-V compound semiconductor, or it can act as a cation to replace Ga as in, e.g., SbN, a material that has been predicted to be metastable under certain conditions.[45] Furthermore, the ionic radius of $Sb^{3-}$ (>207 pm)[46] is much larger than $N^{3-}$ (132 pm), indicating a high degree of local strain during growth. However, the radius of $Sb^{5+}$ (74 pm) matches that of $Ga^{3+}$ (76 pm), implying that Sb should preferably replace Ga due to its better size match (see, e.g., Fig. 1 for a schematic).

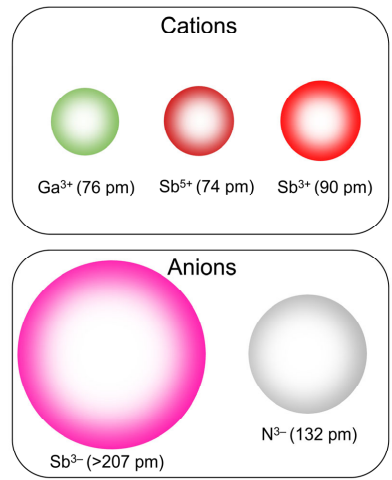

FIG. 1. Schematic representation for the relevant sizes of cations and anions in Sb-doped GaN. Due to the ionic radius match, Sb should preferentially replace Ga as a cation in GaN to avoid large induced strain during growth.

Based on these intuitive arguments, we evaluate the formation energies of Sb-related substitutions in bulk GaN, both as an anion and as a cation, as a function of Fermi level and growth conditions (Fig. 2). We find that Sb replacing Ga is energetically favored for all Fermi energies under N-rich conditions, with isoelectric Sb replacing N ($Sb_N^0$) being favored only under Ga-rich conditions and Fermi levels close to the conduction band (i.e., n-type conditions). Moreover, the preferred charge state of Sb cations for a wide range of Fermi energies is +5 ($Sb^{5+}$, equivalent to $Sb_{Ga}^{2+}$) and switches to +3 ($Sb^{3+}$, equivalent to $Sb_{Ga}^0$) under n-type conditions, i.e., Sb cations act as the deep double donor with an ionization energy of ~420 meV. This specific growth condition has been experimentally reported by Sarney *et al.*[14]; their samples were grown by low-temperature molecular beam epitaxy (MBE) under Ga-rich conditions. Consequently, based on our formation-energy findings, we expect the possible coexistence of $Sb_N^0$, $Sb_{Ga}^0$, and $Sb_{Ga}^{2+}$ in n-type samples with relative concentrations that vary with the growth conditions.

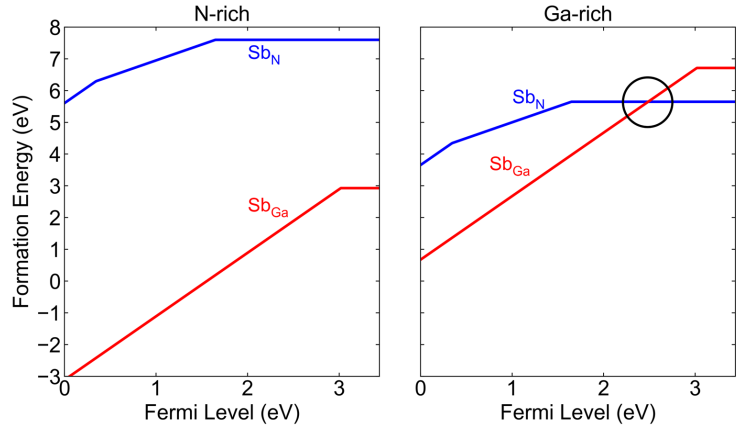

FIG. 2. Formation energies for substitutional Sb on the cation ($Sb_{Ga}$) and anion ($Sb_N$) sites in GaN as a function of Fermi level and growth conditions as calculated with the HSE hybrid functional. The Ga-rich, slightly n-type conditions reported in a previous experimental study[20] correspond to the black circle in the right panel. The comparable formation energies of $Sb_{Ga}$ and $Sb_N$ under Ga-rich n-type conditions reveal the possible coexistence of Sb both as a trivalent anion ($Sb_N^0$, i.e., $Sb^{3-}$ substituting an $N^{3-}$ anion) and as a pentavalent cation ($Sb_{Ga}^{2+}$, i.e., $Sb^{5+}$ replacing a $Ga^{3+}$ cation).

To examine our hypothesis about the possible coexistence of cationic and anionic Sb substitutions in GaN and to explore the formation of different charge states of cationic Sb, it is necessary to quantitatively evaluate further theoretical predictions that can be tested experimentally with, e.g., vibrational spectroscopy. The experimental micro-Raman study of GaSbN epilayers with varying Sb compositions by Chowdhury and Mi,[21] identified two robust additional peaks, labeled $P_{A1}$ and $P_{A2}$ in their reported spectra (these are the only two Sb-related peaks). Although the origins of these modes were attributed to the incorporation of Sb, their connection to Sb-related vibrational modes in GaSbN has not yet been fully established. These additional Raman peaks correspond to phonon modes near Γ, as light scattering can only interact with phonons near the center of the Brillouin zone. We first extract the exact positions of the observed additional peaks from the experimental data in Chowdhury and Mi[21] in Fig. 3 (a). Compared with the Raman spectra of pristine GaN,[47] the $A_1$(LO) mode in experimental GaSbN samples is slightly red-shifted from ~734 $cm^{-1}$ to ~717 $cm^{-1}$, and the two additional peaks emerge as $P_{A1}$ (~658 $cm^{-1}$) and $P_{A2}$ (~470 $cm^{-1}$). Next, we calculate the Raman peaks for the three different Sb substitutions identified in the thermodynamic analysis ($Sb_N^0$, $Sb_{Ga}^0$, and $Sb_{Ga}^{2+}$) as well as for binary GaN. To consider possible interactions between the different substitutions, a co-doped $Sb_N^0$ and $Sb_{Ga}^{2+}$ GaN system is also examined here for completeness. Our calculated Raman peaks for each system, normalized to the strongest peak, are shown and compared to experimental data in Fig. 3 (b). The calculated atomic displacement vectors for the key phonon modes of the various Sb-doped systems are also shown in Fig. 3 (c)-(h). First, our calculated position of $A_1$(LO) peak for pristine GaN agrees well with the experimental reported value at ~735

cm$^{-1}$,[47] while a red shift of the A$_1$(LO) peak is observed in all Sb-substituted systems Fig. 3 (b). We also find that the cationic Sb$_{Ga}^0$ substitution exhibits a strong peak at ~460 cm$^{-1}$, which is close to the experimental P$_{A2}$ peak position. This peak occurs in a gap of the phonon spectra of GaN and corresponds to a localized phonon mode (Fig. 3 (c)) for which the displacements are localized near the Sb atom. At the same time, other Sb substitutions do not give rise to any additional peaks at nearby frequencies. Therefore, a likely origin of peak P$_{A2}$ is the incorporation of Sb$_{Ga}^0$. Additionally, the principal contributions for peak P$_{A1}$ may originate from either Sb$_N^0$ or Sb$_{Ga}^{2+}$, since both of these substitutions exhibit nearby peaks while Sb$_{Ga}^0$ does not give rise to any strong peaks in the range of 600-730 cm$^{-1}$. On the other hand, the additional peaks of both Sb$_{Ga}^{2+}$ and the co-doped Sb$_N^0$ and Sb$_{Ga}^{2+}$ system occur close to the experimental P$_{A1}$ peak. Plus, these two Sb substitutions explain the redshift of the A$_1$(LO) peak observed in experiments. Therefore, the likely origin of peak P$_{A1}$ is Sb$_{Ga}^{2+}$, or the coexistence of Sb$_N^0$ and Sb$_{Ga}^{2+}$, since the presence of Sb$_N^0$ is confirmed by the band-gap reduction in the experimentally grown samples[19,21] as discussed below. Overall, our calculations show that the coexistence of the thermodynamically stable Sb substitutions into GaN (Sb$_N^0$, Sb$_{Ga}^0$, and Sb$_{Ga}^{2+}$) can explain the Raman spectra of Sb-doped GaN.

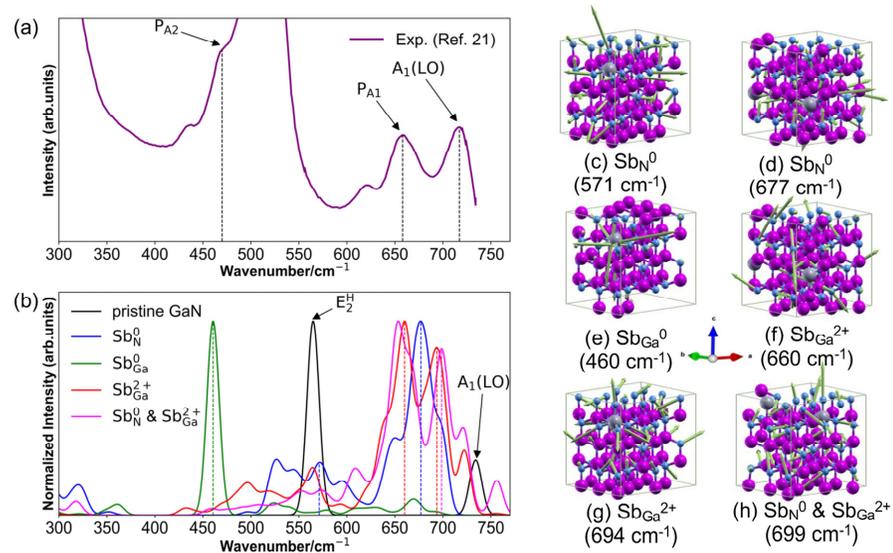

FIG. 3. (a) Partial experimental Raman spectrum (purple line) for Sb-doped GaN extracted from Chowdhury and Mi.[21] Reproduced from *J. Appl. Phys.* 126, 085704 (2019), with the permission of AIP Publishing. (b) Calculated normalized Raman spectra of Sb$_N^0$ (blue line), Sb$_{Ga}^0$ (green line), Sb$_{Ga}^{2+}$ (red line), and Sb$_N^0$ & Sb$_{Ga}^{2+}$ co-doped (pink line) GaSbN in comparison to pristine GaN (black line). All peaks with nonzero intensities are broadened with Gaussian functions with a width of 15 cm$^{-1}$. The origin of peak P$_{A1}$ can be attributed to contributions by both Sb$_N^0$ and Sb$_{Ga}^{2+}$, while P$_{A2}$ is attributed only to Sb$_{Ga}^0$, a cationic substitution that has not been observed through PL measurements. (c) - (h): vibrational modes of the key Raman-active phonons for typical Sb configurations in GaN, the corresponding peak positions are marked with dotted lines in the computed Raman spectra.

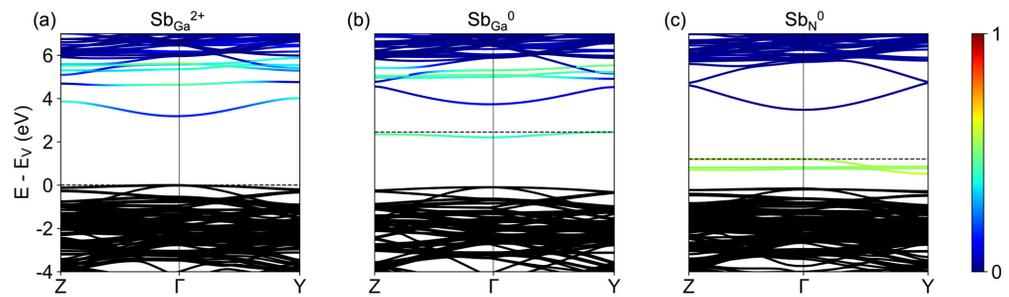

FIG. 4. Band structures of dilute (a) $Sb_{Ga}^{2+}$, (b) $Sb_{Ga}^{0}$, and (c) $Sb_{N}^{0}$ in GaSbN using HSE06, aligned to the top of GaN-contributed valence bands (black). The color bar represents the normalized magnitude of Sb atomic-orbital projections in the conduction bands (blue). The defect bands of $Sb_{Ga}^{2+}$ hybridize with the GaN conduction band edge and create a new hybrid conduction band edge with only ~8.8% band-gap reduction compared to pristine GaN. The band gap of $Sb_{Ga}^{0}$ is ~1.3 eV, which is consistent with the optical absorption result by Sarney *et al.*[14] The band gap of $Sb_{N}^{0}$ is ~2.2 eV, which is consistent with the luminescence experiments by Chowdhury *et al.*[19]

For different incorporations of Sb in GaN, the band gap of the GaN host material can be tuned from the UV (3.4 eV), through the deep visible (2.23-2.41 eV), to the near- and mid-infrared wavelengths (0.8-1.15 eV), which has been experimentally verified.[12,14] The reported peak positions of PL spectra provide direct information on the magnitude of band gaps of various materials. However, the atomic and orbital compositions of the involved bands cannot be determined by experimental means alone, necessitating the use of computational tools to resolve their microscopic origin. By calculating the electronic properties of GaSbN alloy structures, we determine the band structures for various Sb substitutions into GaN such as $Sb_{N}^{0}$ ($Sb^{3-}$), $Sb_{Ga}^{0}$ ($Sb^{3-}$), and $Sb_{Ga}^{2+}$ ($Sb^{5+}$) in GaN. These band structures can fully explain the measurements reported in previous experiments.[10,12,14,19] Figure 4 displays the detailed band structures of $Sb_{N}^{0}$, $Sb_{Ga}^{0}$, and $Sb_{Ga}^{2+}$ GaSbN systems, with the conduction bands that exhibit strong Sb character shown in different colors from blue to red, and the valence bands in black. Based on the analysis of three systems, we derive theoretical explanations for the existing PL spectra of dilute antimonides. As shown in Fig. 4 (c), the $Sb_{N}^{0}$ substitution induces a band gap of ~2.2 eV, consistent with the cathodoluminescence results in Shaw *et al.*[10] and PL results in Chowdhury *et al.*[19] Regarding the electronic configuration of $Sb_{N}^{0}$, anionic Sb is in the 3– charge state just like N, therefore all the 5s and 5p orbitals of Sb are occupied and form a set of three new top valence bands, which can be inferred from the atomic projections of the band structure onto Sb orbitals. Furthermore, Fig. 4 (b) illustrates the effects of $Sb_{Ga}^{0}$ substitution, in which Sb loses its three 5p electrons, and the remaining 5s electrons form a new valence band near the middle of the gap of GaN, resulting in a band gap of ~1.2 eV. Similar band gap values (0.8 - 1.15 eV) have been observed in optical absorption spectroscopy of dilute $GaN_{1-x}Sb_x$ alloys grown in a polar orientation under Ga-rich conditions by Sarney *et al.*[14] In addition, Fig. 4 (a) demonstrates that the band gap of $Sb_{Ga}^{2+}$ is ~3.1 eV as in this case Sb has lost all the electrons in its valence shell, all the empty defect bands are part of the conduction band and hybridize with the original GaN conduction band edge so that the bandgap value is close to that of pristine GaN. Since $Sb_{Ga}^{2+}$ induces only a minor band-gap reduction compared to $Sb_N^0$, the co-existence of $Sb_{Ga}^{2+}$ with $Sb_N^0$ resulting in an ultimate band gap of ~2.0 eV which is still consistent with PL experiments.[14,19] We therefore propose that the co-existence of $Sb_{Ga}^{2+}$ and $Sb_N^0$ cannot be directly inferred via luminescence peaks. Overall, we find that the various thermodynamically stable substitutions of Sb ($Sb_N^0$, $Sb_{Ga}^{2+}$, and $Sb_{Ga}^0$) can consistently explain all experimentally reported optical absorption, PL, and Raman spectra of GaSbN grown under varying conditions.

The selective incorporation of dilute Sb into GaN via controlled growth conditions has significant implications for enhancing device performance, particularly in optoelectronic devices for which drastic band-gap modifications are crucial. This strategy allows for tailored electronic properties, potentially leading to more efficient light-emitting diodes and laser diodes in the long-wavelength part of the visible spectrum. Moreover, previous theoretical studies have uncovered that other group-V elements such as $As^{30,48,49}$ or $Bi^{31,50}$ can also be selectively incorporated into GaN either as cations or anions under varying growth conditions. We, therefore, propose that the dilute incorporation of group V elements into GaN can further broaden the range of band-structure engineering, opening new avenues for dilute group V-doped devices operating across a broader spectrum of wavelengths.

To conclude, we applied DFT calculations to analyze the incorporation of Sb into GaN, and to identify the atomistic origins of the band-gap reduction and the reported features in experimentally measured optical spectra (PL, absorption, Raman). Our thermodynamic calculations show that Sb can incorporate in GaN both as a trivalent anion ($Sb_N^0$) and as a trivalent cation ($Sb_{Ga}^0$) or even a pentavalent cation ($Sb_{Ga}^{2+}$) depending on the growth conditions. Our analysis of the Raman spectra of Sb-doped GaN demonstrates the possible coexistence of $Sb_N^0$ and $Sb_{Ga}^{2+}$ and the presence of $Sb_{Ga}^0$ in GaN. Finally, our band-structure results for different Sb substitutions indicate that $Sb_N^0$ (with the potential coexistence of $Sb_{Ga}^{2+}$) explains the existing PL data, while $Sb_{Ga}^0$ can explain the optical gaps measured in optical absorption experiments. Our work provides insights into the atomistic origins of the electronic and optical properties of dilute antimonide III-nitride semiconductors, paving the way for widely tunable band-structure engineering for applications in long-wavelength light emitters and photocatalysis.

See the supplementary material for the crystal structure of the GaSbN supercell.

**ACKNOWLEDGMENTS**

We thank Zihao Deng, Xiao Zhang, Woncheol Lee, Nick Pant, Amanda Wang, and Kyle Bushick for helpful discussions. This work was supported by the US Army Research Office Award W911NF2110337. Computational resources were provided by the National Energy Research Scientific Computing Center, which is supported by the Office of Science of the U.S. Department of Energy under Contract No. DE-AC02-05CH11231.